\documentclass[11pt]{article}

\usepackage[utf8]{inputenc}
\usepackage[margin=1in]{geometry}
\usepackage{graphicx}
\usepackage{amsmath}
\usepackage{amssymb}
\usepackage{rotating}
\usepackage{graphicx}
\usepackage{wrapfig}
\usepackage{lscape}
\usepackage{rotating}
\usepackage{epstopdf}
\usepackage{float}
\usepackage[nottoc]{tocbibind}
\usepackage{caption}
\usepackage{subcaption}
\usepackage{mathtools}
\usepackage[affil-it]{authblk}
\usepackage{siunitx}
\usepackage[colorlinks=true, allcolors=black]{hyperref}
\usepackage{comment}
\usepackage{cite}

\begin{document}

\title{Efficient numerical frameworks for modelling ultrasonic beams propagating across interfaces}

\author{André Lello de Almeida, Melody Png, Bo Lan\footnote{\protect\label{corresponding_author} \raggedright{\noindent Corresponding authors: bo.lan@imperial.ac.uk}}}
\affil{Department of Mechanical Engineering, Imperial College London, London SW7 2AZ, UK}
\date{\vspace{-5ex}}
\maketitle

\begin{abstract}
Two different frameworks are developed to model the wave field generated by a transducer and propagating through one or more interfaces, and a Quasi-Monte Carlo (QMC) integration scheme is used to numerically evaluate their results. The first method is based on the Rayleigh-Sommerfeld Integral (RSI), further developing a formulation in the literature and improving its capabilities, while the second relies on a high-frequency approximation, using a ray tracing principle. The advantages and limitations of each model are then compared via in-depth investigations on several use cases, culminating in an efficiency and scope assessment. It was found that the RSI-based model performs well if a large number of field points is needed, such as when modelling a full image of the field. Conversely, for a large number of interfaces, such as when modelling the field through a thin-layered material, the most efficient model was the ray tracing formulation, since it was unnecessary to propagate the field between all the interfaces first. This was especially noticeable for applications requiring only the evaluation of the field at a few points on the other side of multiple interfaces.
\end{abstract}
\noindent \textbf{Keywords:} propagation through interfaces; wave field diffraction; quasi-Monte Carlo method.

\section{Introduction}
\label{sec,intro}
Ultrasonic Non-Destructive Evaluation (NDE) relies heavily on understanding how wave fields propagate from transducers to the evaluated media. Immersion systems, which is the most widely employed ultrasonic NDE configuration, where the sample to be examined is immersed in a liquid, or dry coupled inspections, are prime examples of the intrinsic relation between the acquisition of experimental parameters and the way transducer generated waves interact with the propagation medium, and liquid-solid or solid-solid interfaces. For that reason, the modelling of ultrasonic wave fields through interfaces plays an essential role in predicting and understanding the extent to which beam propagation influences the results of experiments performed with transducers \cite{Schmerr_2016}.

Despite the practical importance of these models, evaluating wave fields generated by transducers remains amongst the most complex tasks when analysing ultrasonic systems \cite{Schmerr_2016}. This complexity arises from the fact that these waves are not perfectly plane, being altered by a superposition of edge terms responsible for all beam diffraction effects. All of this contributes to the subject still being an active field of research \cite{Banerjee et al_2007,Zhang et al_2021}, regardless of the existence of an already very extensive literature.

It is possible to separate existing transducer beam models into three categories: analytical, semi-analytical, and numerical methodologies \cite{Zhang et al_2021}. Analytical methods offer highly accurate predictions, but lack flexibility in the sense that they are not generally applicable to most transducer and interface geometries \cite{Zhang et al_2021}. Only for simple cases, such as a circular piston radiating into a fluid \cite{Williams_1951,Rogers and Van Buren_1974}, do we have readily available formulae.

On the other hand, semi-analytical methodologies attempt to address this issue by assuming certain approximations when solving the analytical equations. This usually results in poorer agreement with the exact solution, but allows them to maintain the speed of analytical methods while acquiring a much broader spectrum of applicability \cite{Schmerr_2016}. Their attractiveness remains, however, for problems that require three-dimensional modelling, as these introduce computational complexities that, in the case of element-based numerical methods, will lead to very large models. As examples of this category, we have the paraxial approximation \cite{Schmerr et al_1995,Lerch et al_1995} and multi-Gaussian beam models \cite{Schmerr_2016,Wen and Breazeale_1988,Kim et al_2006}.

In the last group mentioned, we can include numerical implementations of integral theorems from diffraction theory (e.g. the Rayleigh-Sommerfeld Integral (RSI) \cite{Born and Wolf_1980,Goodman_2017}), the widely used Finite Element Method (FEM) \cite{Lord et al_1988,Wojcik et al_1993,Bathe_2014}, the finite differences method \cite{Alterman and Karal_1968,Scandrett and Achenbach_1987,Alia et al_2004}, the boundary element method \cite{Niwa et al_1986,Aldrin et al_2001}, and the distributed point source method \cite{Banerjee et al_2007,Placko and Kundu_2001}. These methods are regarded as highly accurate, often comparable to the exact solutions, but in turn requiring longer computation times to achieve the desired outcome \cite{Zhang et al_2021}. This is especially true for three-dimensional problems where the number of elements tends to be significantly higher, particularly when fluids are involved. Despite this, the fact that these models provide the best accuracy for the largest range of applied problems constitutes the main motivator for continued research on their improvement \cite{Banerjee et al_2007,Lhemery et al_1999,Liu et al_2011,Huthwaite_2014}.

Notwithstanding the number of available models, choosing any of them requires a compromise between computation time, accuracy, and applicability, turning the general question of model suitability into an application dependent one. With that in mind, an inspiring previous work has showed that using a Quasi-Monte Carlo (QMC) integration scheme reduces the time consumption of some numerical algorithms, while maintaining a similar level of accuracy expected from the most precise methods \cite{Zhang et al_2021,Zhang et al_2022}. For example, Zhang et al use a QMC implementation of the RSI to calculate the wave field generated by a transducer radiating into a fluid \cite{Zhang et al_2021} and interacting with a boundary \cite{Zhang et al_2022}, resulting in a reported increase in speed of one to two orders of magnitude. This increase is especially important for modelling the intrinsically three-dimensional wave fields generated by finite transducers, which are routinely approximated as two-dimensional in an attempt to decrease the computational load when applying other numerical methods, such as FEM.

Besides this, another interesting aspect of their work is that the QMC method is used only at the end of the algorithm, facilitating the evaluation of certain integrations required for the computation of the field. This demonstrates that the technique can be used as a generic numerical means for assessing complex integrations, and is consequently applicable to other integral-based numerical models. Naturally, this prompts the question of which numerical method benefits the most from the use of a QMC integration scheme, instead of the more widespread meshing procedures.

Therefore, in this paper we offer two different frameworks for modelling wave fields, in order to solve the general three-dimensional problem of a transducer beam interacting with multiple interfaces in succession. The first one is based on the work of Zhang et al \cite{Zhang et al_2021,Zhang et al_2022}, building on its foundations and providing important improvements to their solution, while the second implementation uses a ray tracing framework, inspired by variational principles in optics.

In particular, the proposed version of ray tracing is able to correctly predict the intensity of the field everywhere, even when focusing is present, which makes it suitable for extending its capabilities to more complex materials, like anisotropic ones. While the same should be possible for the RSI-based model, it is expected to be not as straightforward and more time consuming to do so. Despite these differences, evaluating both of these formulations with the proposed QMC integration scheme is shown to bring massive gains in efficiency when compared to their regularly meshed versions or other precision methods. As a result, by using both suggested methodologies, we hope to pave the way for further improvements in our computer models, enabling them to solve increasingly complex problems in the shortest time possible.

Results for several cases are presented, namely for an unfocused transducer at oblique incidence, a focused transducer at normal incidence, and a multi-layered material being inspected at oblique incidence by an unfocused transducer. This broad range of use cases was selected to show the general applicability of both methods, as well as reveal their inherent merits and shortcomings. A discussion on the computational efficiency and scope of both implementations follows from this, concluding with some general remarks about when to use each methodology and possible applications.

\section{Theoretical Framework}
\label{sec,theoretical_framework}
As mentioned previously, the complexity of the problem at hand rests not only on correctly representing the complicated three-dimensional field created by a finite-sized transducer, but also in understanding how this wave interacts with a boundary, being both part reflected and part refracted. To address this issue, many methods use the superposition principle, allowing them to model the complex beam as a sum of simpler, better understood waves \cite{Rayleigh_1945,King_1934,Schoch_1941}. This also constitutes the reason why a large portion of numerical methods rely on the evaluation of integrations, as a means to represent these sums.

The fact that the QMC method is needed only to evaluate such operations in our chosen numerical model, opens up several possible avenues of investigation regarding which implementation is better overall or better for specific applications. In fact, starting from the RSI, three different types of approaches could be taken to evaluate the wave field generated by a transducer and interacting with boundaries: a multiple surface integral model \cite{Harris_1981}, using the unchanged RSI, a single surface integral model \cite{Nayfeh and Chimenti_1984,Pialucha_1992}, employing an angular spectrum of plane waves approach, or a line integral model, based on the idea of boundary diffraction waves \cite{Rubinowicz_1917,Marchand and Wolf_1962}.

Although the merits of the boundary diffraction wave approach are undeniable, in terms of the physical insight it gives into the mechanisms responsible for diffraction \cite{Marchand and Wolf_1962}, adapting it to more complex boundaries is often a difficult, even if possible, task. The line integral resulting from the edge wave term becomes mathematically intractable for cases other than normal incidence onto a planar boundary, making this approach less fruitful for practical modelling \cite{Schmerr_2016}. Consequently, this section will focus exclusively on developing multiple and single surface integral formulations.

\subsection{Rayleigh-Sommerfeld diffraction}
\label{sec,rayleigh_diffraction}
Diffraction by an aperture in a planar opaque screen is the classical example used as an introduction to scalar diffraction theory \cite{Born and Wolf_1980,Goodman_2017}. However, the equations are easily adapted to the case of an infinite planar semi-transparent screen, which is exactly what a boundary between two media with finite impedance is.

Consider the boundary separating two different media shown in Figure \ref{fig,kirchhoff-helmholtz_integral_theorem}. A source $P$ in the first medium is generating an acoustic field that transmits through the interface and causes an observable effect at the observation point $P_o$. We can surround the observation point by two intersecting surfaces: one along the interface, $S_1$, and a spherical surface of radius $R$ centered at the observation point, $S_2$. These two surfaces form an enclosed volume $V$, making the field inside dependant only upon its value at the boundaries \cite{Goodman_2017}. This relation can be explicitly stated using the Kirchhoff-Helmholtz integral theorem,

\begin{equation}
\label{eqn,helmholtz_kirchhoff}
\varphi\left(P_o\right) = \iint_S \left(\varphi\frac{\partial G}{\partial \boldsymbol{n}} - G\frac{\partial\varphi}{\partial \boldsymbol{n}}\right) dS
\end{equation}

\noindent where $\varphi$ represents the field at a particular point and $G$ is the Green's function at the observation point generated by an impulse at a boundary point. It is important to note that, contrary to the literature \cite{Born and Wolf_1980,Goodman_2017}, in this paper we are assuming that the normal to the surface, $\boldsymbol{n}$, is pointing inwards, and that vector $\boldsymbol{r}$ starts at an interface point, $P_I$, and ends at $P_o$. This change strives to maintain a conceptually cleaner picture of the physical mechanisms behind the propagation of spherical waves, wherein interface points act as sources generating the field at the observation point, instead of the reverse scenario.

To solve for the field at the observation point we only need to evaluate the integrand of the previous equations at both surfaces. It is clear that the shape or size of $V$ will not have an effect on the field inside, therefore we can simply increase the value of $R$ arbitrarily. By making the radius approach infinity, we can effectively eliminate the contribution of surface $S_2$ out of the integral, due to the Sommerfeld radiation condition \cite{Goodman_2017}, provided that only outgoing waves exist.

Of course, for this argument to be consistent, one would need to change our convention and consider the Green's functions as being generated by the observation point, because only then can we say that purely outgoing waves are present. However, since the value of $G$ depends exclusively on the distance between points, it does not matter which one is the source and which is the receiver, allowing us to change the convention defined in Figure \ref{fig,kirchhoff-helmholtz_integral_theorem} momentarily and use an argument based on the radiation condition.

\begin{figure}[!t]
    \centering
    \includegraphics[width=0.35\linewidth]{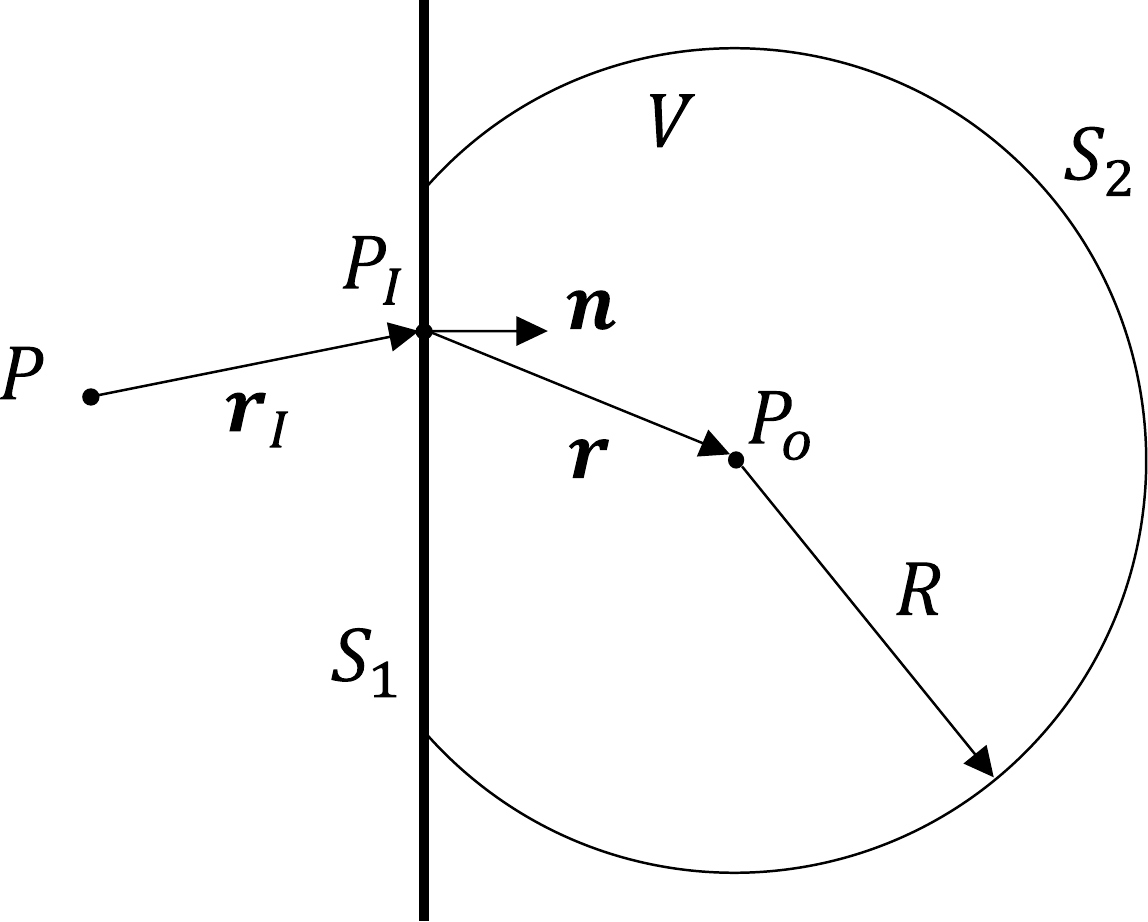}
    \caption{Rayleigh-Sommerfeld diffraction by a semi-transparent plane interface. The total field at an observation point $P_o$ inside the enclosed volume $V$ depends only on the value of the field in the surrounding surfaces $S_1$ and $S_2$. This figure represents a two-dimensional section of a volume.}
    \label{fig,kirchhoff-helmholtz_integral_theorem}
\end{figure}

As a consequence, only the interface $S_1$ contributes to the field at $P_o$ and the first Rayleigh-Sommerfeld solution \cite{Goodman_2017}, henceforth designated by the RSI, can be applied directly,

\begin{equation}
\label{eqn,rayleigh_sommerfeld}
\varphi\left(P_o\right) = -\frac{i}{\lambda} \iint_S \varphi\frac{e^{ikr}}{r} \left(\frac{1}{ikr} - 1\right)\cos{\theta} dS
\end{equation}

\noindent where $\lambda$ and $k$ are, respectively, the wavelength and wavenumber of the propagating waves, $r$ is the distance between $P_I$ and $P_o$, and $\theta$ is the angle between $\boldsymbol{n}$ and $\boldsymbol{r}$. The first solution is often preferred to the second Rayleigh-Sommerfeld solution or the Fresnel-Kirchhoff diffraction formula due to its simplicity, since it involves only the amplitude of the field and not its first derivative.

In the previous equation, it is apparent that the interface functions as a collection of point sources, agreeing with the general picture of wave propagation drawn by the Huygens' principle. However, as others have pointed out before \cite{Miller_1991}, those sources behave instead as dipoles and their radiation profile is modulated by a sinusoidal directivity pattern, unlike their usually considered omnidirectionality. This stands in contrast to the formulation used in Zhang et al \cite{Zhang et al_2022}, where those two specific behaviours were not considered.

One could initially imagine that the dipole behaviour is unimportant for far field applications, therefore turning the monopole assumption into a sensible simplification. However, for layered media with relatively thin layers (around a few wavelengths), the dipole effect might remain significant and not considering its influence can create erroneous results. Besides this, the dipole term constitutes the main component of the amplitude near the singularity, making it a fundamental part when assessing the wave field compliance with the imposed boundary conditions. This is a very important point to consider when attempting a validation of any formulation, since any violation of the boundary conditions over the interface will create a physically invalid model.

\begin{figure}[!t]
    \centering
    \includegraphics[width=0.9\linewidth]{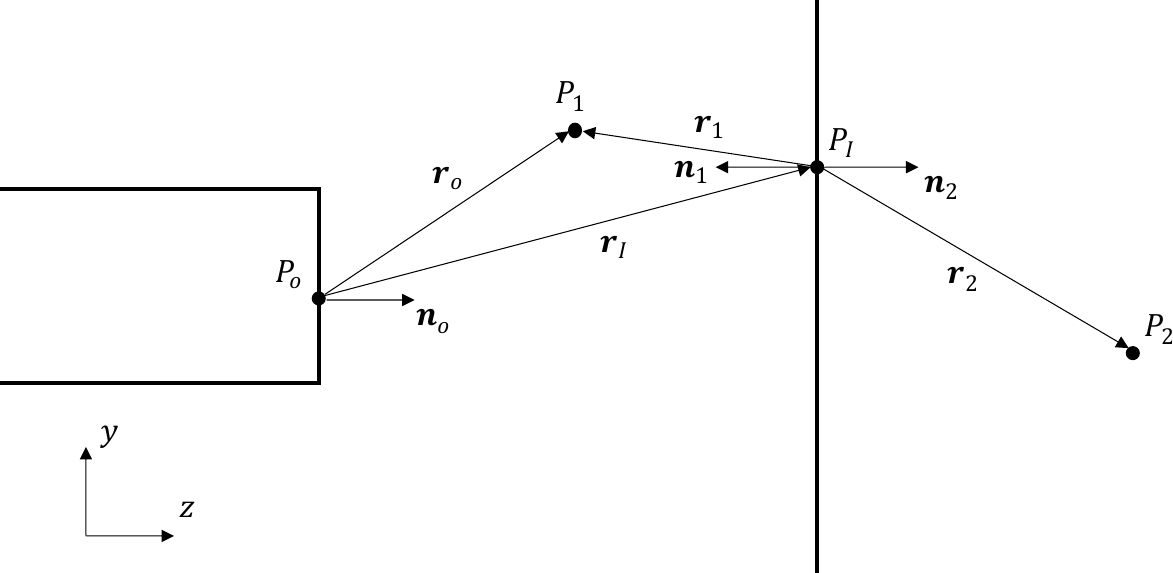}
    \caption{Modeling of an acoustic field through an interface using the RSI formulation. The total field in the first medium is the result of a direct contribution from transducer sources (incident field) and interface sources (reflected field), while the field in the second medium is due only to interface sources (transmitted field).}
    \label{fig,rayleigh-sommerfeld_integral}
\end{figure}

As for the directivity function, its absence will result in the overestimation of the field strength on the other side of the interface, since every point source concentrates most of its emitted energy in the forward direction, as opposed to the omnidirectional emission of a point source in free space. In the work of Zhang et al \cite{Zhang et al_2022} these two effects were not fully taken account of, and some inaccuracies in the calculated field emerge upon close inspection: the pressure fields of their solutions are not consistent across the interface and the transmitted wave fields are stronger compared to FE simulations. Our modifications to the existing model will ensure the recovery of the full accuracy of the solution, as will be shown later in Section \ref{sec,boundary_conditions}.

Following this, consider a transducer sending acoustic waves through an interface between two different media, as in Figure \ref{fig,rayleigh-sommerfeld_integral}. The total acoustic field in the first medium is the result of two components, incident and reflected, while in the second only one exists, transmitted. The incident field is calculated using the direct contribution from all points in the face of the transducer, $P_o$, onto a particular field point, $P_1$. That being so, we can calculate the incident field using a cylindrical piston transducer as the source \cite{Seki et al_1956},

\begin{equation}
\label{eqn,incident_field}
\varphi_i\left(P_1\right) = -\frac{v_o}{2\pi} \iint_S \frac{e^{ik_1r_o}}{r_o} dS
\end{equation}

\noindent where $v_o$ is the velocity amplitude in the z-direction on the face of the transducer, $k_1$ is the wavenumber in medium 1, and $r_o$ is the distance between a point in the face of the transducer $P_o$ and a general point in the first medium $P_1$.

At first, it might seem that we are neglecting the dipole and directivity effect in the previous equation after all the arguments given in favour of their use. However, we just chose to use the second Rayleigh-Sommerfeld solution in this case, which involves the Green's function directly and the derivative of the field value along the normal direction to the surface, or, in other words, the velocity along the direction perpendicular to the face of the transducer $v_o$.

For the reflected and transmitted fields, it is possible to proceed similarly if one resorts to considering all the interface points as sources themselves, via the RSI formulation,

\begin{equation}
\label{eqn,exact_reflected_field}
\varphi_r\left(P_1\right) = \frac{i}{\lambda_1} \iint_S \left(\varphi_i + \varphi_r\right)\frac{e^{ik_1r_1}}{r_1} \left(\frac{1}{ik_1r_1} - 1\right)\cos{\theta_1} dS
\end{equation}

\begin{equation}
\label{eqn,exact_transmitted_field}
\varphi_t\left(P_2\right) = -\frac{i}{\lambda_2} \iint_S \frac{\rho_1}{\rho_2}\left(\varphi_i + \varphi_r\right)\frac{e^{ik_2r_2}}{r_2} \left(\frac{1}{ik_2r_2} - 1\right)\cos{\theta_2} dS
\end{equation}

\noindent where $\rho$ represents the density and the index refers to which medium each variable relates to. It is pertinent to observe the fact that one of the boundary conditions has already been incorporated into our formulation in equation (\ref{eqn,exact_transmitted_field}). This is because, in order to satisfy continuity of pressure (or normal stress), the amplitude of the total field on each side of the boundary should be mediated by the ratio of material densities,

\begin{equation}
\label{eqn,boundary_condition}
\rho_1 \left[\varphi_i\left(P_I\right) + \varphi_r\left(P_I\right)\right] = \rho_2 \varphi_t\left(P_I\right)
\end{equation}

\noindent for all interface points.

According to equations (\ref{eqn,exact_reflected_field}) and (\ref{eqn,exact_transmitted_field}), in order to calculate the reflected and transmitted fields, we must first evaluate the total field at the interface. However, this is not possible without previous knowledge about the reflected field. Therefore, those equations are not solvable in their current implicit form, without the use of iterative methods.

To address this issue, it is possible to introduce certain assumptions regarding the value of the reflected field at the interface. A natural approximation would be to consider that the reflected field is proportional to the incident field and that the constant of proportionality is given by the plane wave reflection coefficient. This will eliminate the possibility of accounting for certain phenomena, like the propagation of interface waves, but away from the boundary the transmitted field should still be representative.

As a final result, and using the approximation mentioned, it becomes possible to solve for the transmitted field using the RSI formulation,

\begin{equation}
\label{eqn,approximated_transmitted_field}
\varphi_t\left(P_2\right) = -\frac{i}{\lambda_2} \iint_S \frac{\rho_1}{\rho_2}\left(1 + R_{12}\right)\varphi_i\frac{e^{ik_2r_2}}{r_2} \left(\frac{1}{ik_2r_2} - 1\right)\cos{\theta_2} dS
\end{equation}

\noindent with $R_{12}$ representing the plane wave reflection coefficient for a particular transducer inclination,

\begin{equation}
\label{eqn,reflection_coefficient}
R_{12} = \frac{\rho_2c_2\cos{\alpha_1} - \rho_1c_1\cos{\alpha_2}}{\rho_2c_2\cos{\alpha_1} + \rho_1c_1\cos{\alpha_2}}
\end{equation}

\noindent where $c$ represents the wave velocity in each respective medium, $\alpha_1$ is the angle of the incident beam, and $\alpha_2$ is the angle of the refracted beam.

\subsection{Ray tracing}
\label{sec,ray_tracing}
When interfaces are included, the RSI formulation requires that we solve a certain number of nested surface integrals to obtain the transmitted field, hence the name of a multiple surface integral model. However, it is possible to reduce the RSI to a single surface integral model, by restricting the domain of allowed wavenumbers, such that only high frequency waves persist. The Kirchhoff approximation is particularly useful in this context, since any sufficiently high frequency wave will behave as a plane wave when reflecting or refracting through an interface, simplifying the calculations necessary to define the reflection and transmission coefficients \cite{Schmerr_2016}.

With this in mind, we start by expanding the RSI for the incident field into an angular spectrum of plane waves \cite{Schmerr_2016},

\begin{equation}
\label{eqn,incident_angular_spectrum}
\varphi_i\left(P_1\right) = -\frac{iv_o}{4\pi^2} \iint_S \int_{-\infty}^{\infty} \int_{-\infty}^{\infty} \frac{e^{i\boldsymbol{k}_1\cdot\boldsymbol{r}_o}}{k_{1z}} dk_x dk_y dS
\end{equation}

\noindent borrowing from the notation used in Figure \ref{fig,rayleigh-sommerfeld_integral}.

To calculate the transmitted field resulting from this excitation one simply needs to transmit each individual plane wave, travel along a specific direction, independently of all the others, discarding the need for additional assumptions like in the RSI formulation. Therefore, the transmitted field is \cite{Schmerr_2016},

\begin{equation}
\label{eqn,transmitted_angular_spectrum}
\varphi_t\left(P_2\right) = -\frac{iv_o}{4\pi^2} \iint_S \int_{-\infty}^{\infty} \int_{-\infty}^{\infty} T_{12} \frac{e^{i\left[k_x\left(x_2-x_o\right) + k_y\left(y_2-y_o\right) + k_{1z}\left(z_I-z_o\right) + k_{2z}\left(z_2-z_I\right)\right]}}{k_{1z}} dk_x dk_y dS
\end{equation}

\noindent with $T_{12}$ representing the plane wave transmission coefficient,

\begin{equation}
\label{eqn,transmission_coefficient}
T_{12} = \frac{2\rho_1c_2\cos{\theta_1}}{\rho_2c_2\cos{\theta_1} + \rho_1c_1\cos{\theta_2}}
\end{equation}

An important detail to note is the fact that the wavenumber components parallel to the interface, $k_x$ and $k_y$, are conserved, which is a direct result of the application of Snell's law. This defining feature is one of the main reasons why the angular spectrum formulation can be applied to fields transmitted through interfaces as directly as in equation (\ref{eqn,transmitted_angular_spectrum}). In fact, adding extra interfaces to our system is as trivial as calculating the right phase difference to a point in a certain medium and multiplying the integrand by its respective transmission coefficient. This stands in stark contrast with the RSI formulation, where every new interface introduces a new nested surface integral.

The integral representing the transmitted field generated by the transducer, as described above, is still an exact model of the system, entirely equivalent to the RSI formulation. However, in order to evaluate the integration over all wavenumber directions, further assumptions are necessary. If we consider that all plane waves are high frequency ones, the method of the stationary phase can be applied to solve the integration over $k_x$ and $k_y$ analytically.

Thus, the angular spectrum integral is transformed into a ray tracing method, where all plane waves can be described by equivalent acoustic rays traveling in the same direction as the wavenumber vector. The transmitted field becomes \cite{Schmerr_2016},

\begin{equation}
\label{eqn,transmitted_ray_tracing}
\varphi_t\left(P_2\right) = -\frac{v_o}{2\pi} \iint_S T_{12} \frac{e^{i\left(k_1D_1 + k_2D_2\right)}}{\sqrt{\left(D_1 + \frac{c_2\cos{\theta_1}^2}{c_1\cos{\theta_2}^2}D_2\right)\left(D_1 + \frac{c_2}{c_1}D_2\right)}} dS
\end{equation}

\noindent where $D_1$ and $D_2$ are the total distance travelled in each respective medium.

The problem now consists of determining the geometrical laws that define the ray tracing equations, so that Snell's law is satisfied for every pair of incident and refracted rays. Instead of analysing the geometry directly, we can use Fermat's principle of least time, which states that a ray in going from point A to point B must traverse a path length which makes the total time of flight stationary \cite{Born and Wolf_1980},

\begin{equation}
\label{eqn,fermat_principle}
\delta \int_A^B \frac{ds}{c} = 0
\end{equation}

\noindent with $c$ being the wave speed for each particular point in the ray path.

The merits of using a variational principle to solve our geometrical problem are evident when considering more complex interfaces, such as curved boundaries or inhomogeneous propagation media. Describing the mechanics of our system using functionals also has the opportune advantage of letting the symmetries, and corresponding conserved quantities, emerge as natural properties. This will be particularly useful for the algorithmic implementation of the ray tracing formulation.

For two isotropic media separated by a planar boundary, it is easy to show that Fermat's principle leads to the following path from the transducer to a point in medium 2,

\begin{equation}
\label{eqn,ray_tracing}
\sqrt{\left(x_2-x_o\right)^2 + \left(y_2-y_o\right)^2} = \frac{C\left(z_I-z_o\right)}{\sqrt{S_1^2 - C^2}} + \frac{C\left(z_2-z_I\right)}{\sqrt{S_2^2 - C^2}}
\end{equation}

\noindent where $C$ is the conserved quantity and $S$ represents the slowness for each particular medium.

Therefore, for any two endpoints it is possible to calculate the path taken by an acoustic ray using equation (\ref{eqn,ray_tracing}), where all the information necessary to determine the angles and distances travelled by each particular ray is encoded in the unknown conserved quantity. That being so, to effectively solve the equation in terms of $C$, a root-searching algorithm needs to be employed, except on two very special cases. If there is a single propagation medium, the equation is trivial, because the ray will travel in a straight line. Also, for two different isotropic media, one can rearrange the ray tracing equation into a fourth-order polynomial equation, which can be solved analytically using Ferrari's method \cite{Weston et al_2012}.

It is pertinent to comment on the necessity of ray tracing, with its requirement for a root-finding algorithm, since our solution does not seem to benefit from a uniform spatial discretisation at the destination. One could instead solve the space-coordinate integration of equation (\ref{eqn,transmitted_angular_spectrum}) first by decomposing the incident signal into its spatial frequency components. In fact, this methodology has been used in the literature \cite{Pialucha_1992} for modelling transducer fields interacting with multi-layered plates, albeit in two dimensions.

This Fourier decomposition approach will create a uniform angular discretisation at the source, with every single plane wave propagating simultaneously to all field points, waiving the need to carry out rays through specific paths. At first glance this seems like a more efficient solution to our model, since there is no time spent on finding ray paths from source to receiver. However, since the spatial frequency integration relies on the use of a set of random wavenumbers, to be able to accurately represent the frequency domain amplitude, interpolation must be performed. The decrease in computation time due to not including a root-searching algorithm is directly replaced by a similar increase for interpolating the complex amplitude of each random plane wave, rendering the Fourier decomposition method and the ray tracing equivalent in terms of computational efficiency.

Finally, one might ask what the meaning of the conserved quantity $C$ is and how to relate it with the orientation of the travelling rays. However, by recalling that Snell's law must be satisfied for every possible ray path, $C$ must surely be,

\begin{equation}
\label{eqn,root_searching}
C = \frac{\sin{\theta_1}}{c_1} = \frac{\sin{\theta_2}}{c_2}
\end{equation}

\subsection{Quasi-Monte Carlo method}
\label{sec,monte_carlo}
As mentioned previously, the QMC method is used only to evaluate the surface integral at the end of each formulation, which means it can be directly applied without any model-specific modifications. To evaluate a general multi-dimensional integral over a s-dimensional unit cube $I^s$, we can approximate it by the average of a sample of pseudo-random points \cite{Morokoff and Caflisch_1995},

\begin{equation}
\label{eqn,monte_carlo}
\int_{I^s} f(\boldsymbol{x}) d\boldsymbol{x} \approx \frac{1}{N} \sum^N_{i=1} f(\boldsymbol{x}_i)
\end{equation}

\noindent with every integration point $\boldsymbol{x}_i$ belonging to the integration domain $I^s$.

For our purpose, this means that instead of using a discretisation scheme to evaluate an integral over a certain domain, be it the face of a transducer or an interface, we assign a large number of random sources to the area of interest and compute the average value of the field resulting from their superposition. Of course, this means that the sampling process will be different in each formulation, as the change in integration domain is accommodated. For the RSI-based model, using equation (\ref{eqn,approximated_transmitted_field}), both the face of the transducer and all the interfaces are populated by a pseudo-random set of points, which serves as the integration sample. On the other hand, in the ray tracing formulation, using equation (\ref{eqn,transmitted_ray_tracing}), only the transducer is subjected to the pseudo-random sampling.

To generate all the necessary pseudo-random samples the Halton sequence is used, due to its low discrepancy properties and higher convergence rates for lower-dimensional problems, when compared to other pseudo-random sequences. This higher convergence rate guarantees that with an increase in the number of points, the approximation will converge monotonically to the exact value of the integral as efficiently as possible \cite{Morokoff and Caflisch_1995}.

For every interface, the sample area is considered to be a square with sides equal to three times the size of the transducer's diameter, which means that sampling a uniform collection of points is trivial,

\begin{equation}
\label{eqn,square_sampling}
x_I = 3a\left[2H(n)-1\right] \,\,\,\,\, , \,\,\,\,\, y_I = 3a\left[2H(n)-1\right]
\end{equation}

\noindent where $a$ is the radius of the transducer and $H(n)$ is the Halton sequence generated by the prime number $n$. Ideally, an infinite interface would need to be sampled over its full extent, but because the field decays significantly when observing points far from the central axis of the beam, a sampling distance of $3a$ was deemed sufficient to accurately represent it \cite{Zhang et al_2022}.

As for the face of the transducer, since it is a circle, sampling a uniform group of points in the radial and polar directions requires a correction to the distribution,

\begin{equation}
\label{eqn,circle_sampling}
r_o = a\sqrt{H(n)} \,\,\,\,\, , \,\,\,\,\, \theta_o = 2\pi H(n)
\end{equation}

\section{Numerical Results}
\label{sec,numerical_results}
Several different numerical examples are used in this section to provide a basis of comparison between the two formulations established previously. All acoustic fields are generated by a single piston transducer either at normal or oblique incidence and every numerical case consists of at least one liquid-solid interface.

\begin{figure}[!t]
    \centering
    \includegraphics[width=1\linewidth]{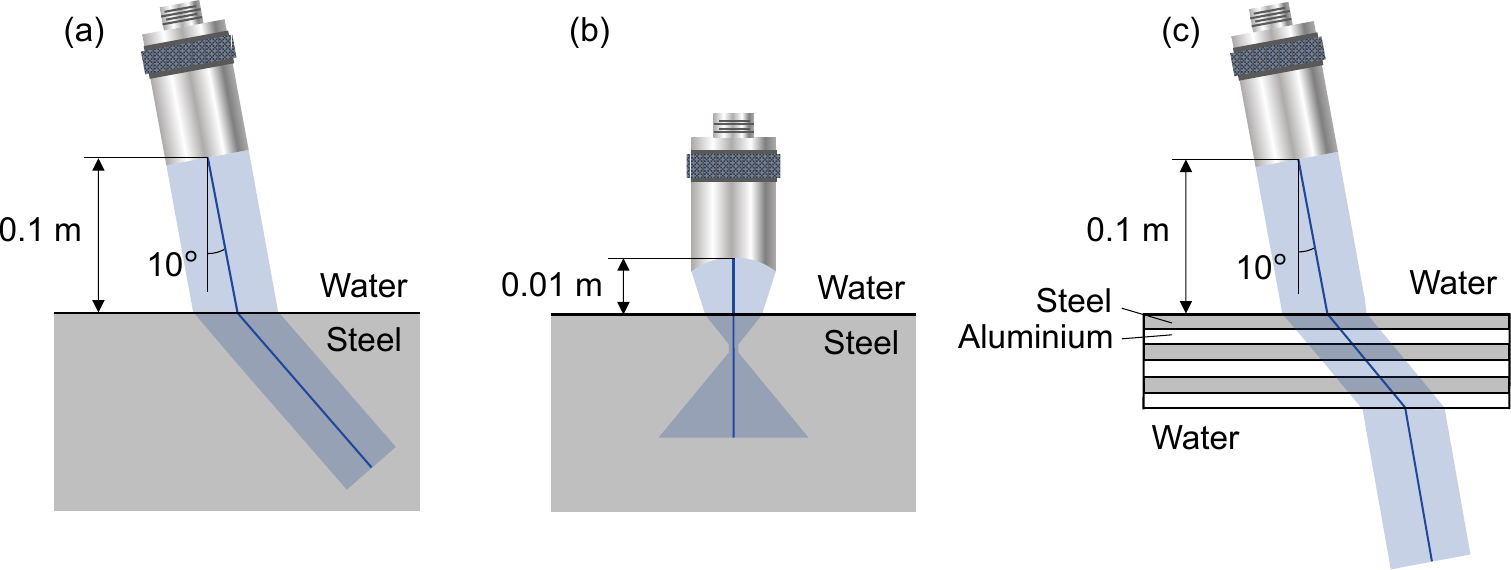}
    \caption{Representation of the geometry used for the system in each particular study case: oblique incidence (a), focused transducer (b), and multi-layered medium (c).}
    \label{fig,transducer_system}
\end{figure}

A frequency of $f=2.5$ MHz and a radius of $a=6.35$ mm are used as transducer parameters and, for the first two numerical cases, only two materials are present: water ($\rho_1=1000$ kg/m\textsuperscript{3} and $c_1=1480$ m/s) and steel ($\rho_2=7800$ kg/m\textsuperscript{3} and $c_2=5840$ m/s). In the very last example, a third material is present, in the form of aluminium ($\rho_3=2700$ kg/m\textsuperscript{3} and $c_3=6409$ m/s). It should be noted that the shear velocity is not being used, as every material is considered a purely acoustic medium. In the first and third cases, the water path is set to be $L=0.1$ m, in order to allow the full development of the beam, and avoid near-field complications, while for the focused transducer, the water path is set to $L=0.01$ m. To enable a fairer comparison between both methods, a consistent number of sample points, $N=10^5$, is used for every surface integral. As a final note, all field simulations were three-dimensional, but, to avoid overly complex plots, a single two-dimensional slice of the field was considered, consisting of a grid of 401x401 points in the yOz plane. For a detailed representation of the geometry of each system, refer to Figure \ref{fig,transducer_system}.

\subsection{Oblique incidence}
\label{sec,oblique_incidence}
For a water-steel system with a transducer at a $10^\circ$ angle, the simulation results are plotted in Figure \ref{fig,oblique_incidence}. Both methodologies are employed to represent a slice of the three-dimensional wave field for $x=0$, the on-axis amplitude, and the off-axis distribution at $z=0.12$ m. The off-axis amplitude is being plotted with the central beam being the reference for the horizontal axis. This can be observed by the red dashed lines in Figures \ref{fig,oblique_incidence}a and \ref{fig,oblique_incidence}b, representing the on-axis and off-axis coordinates of the central axis of the beam.

\begin{figure}[!t]
    \centering
    \includegraphics[width=1\linewidth]{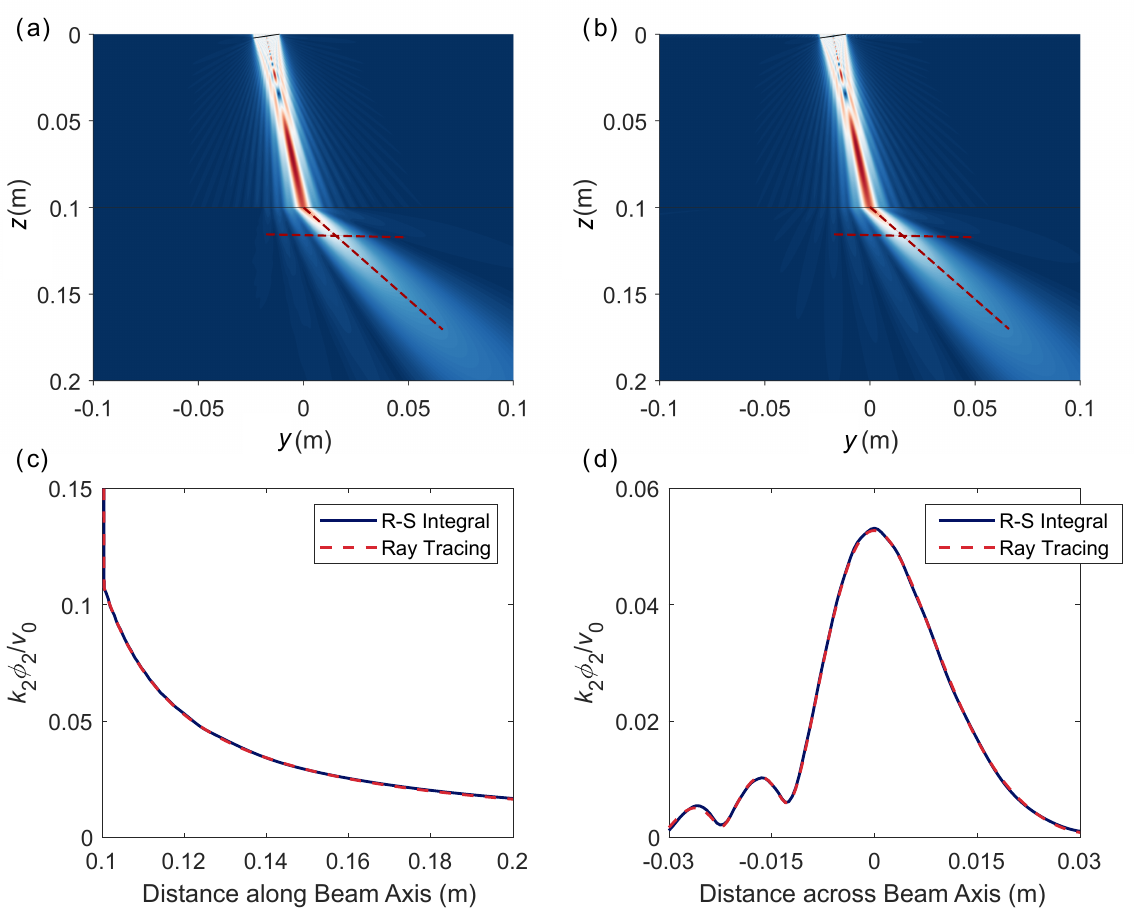}
    \caption{Wave field generated by a transducer at oblique incidence and transmission through one interface. The full wave field for $x=0$ is modelled using the RSI (a) and the ray tracing formulation (b). Detailed versions of the field are portrayed using the on-axis amplitude of the transmitted field (c) and the off-axis amplitude distribution at $z=0.12$ m (d) for both formulations. The red dashed lines represent the on-axis and off-axis coordinates of the central axis of the beam. The intensity of the field is represented using a normalised velocity amplitude in each medium, $k_i\phi_i/v_o$.}
    \label{fig,oblique_incidence}
\end{figure}

It is noteworthy to explain that because of the impedance change from water to steel the transmitted field is much weaker than the incident one. Therefore, to grant a better visualisation of the field in Figures \ref{fig,oblique_incidence}a and \ref{fig,oblique_incidence}b, the impedance of steel was matched to that of water. This was performed by adjusting the transmission coefficient, so that it assumes a value of 1, or equivalently by matching the densities of the two media, in accordance to the literature \cite{Zhang et al_2022}.

Good agreement is obtained between the methods, both near and away from the central axis of the beam. The fact that the predicted wave fields match so consistently should serve as a good indication of the validity of these methodologies. Nevertheless, to further prove this point, a study of the behaviour of these wave fields near the interface is shown in the next subsection.

\subsection{Boundary conditions}
\label{sec,boundary_conditions}
As stated in section \ref{sec,rayleigh_diffraction}, we believe that the assumptions made in the work of Zhang et al \cite{Zhang et al_2022} lead to the pressure field being discontinuous across the liquid-solid interface, as their resulting wave fields do not satisfy the prerequisite boundary conditions. To prove this point and, at the same time, provide further validation for our own models, additional investigation into the behaviour of these solutions near the interface was deemed necessary.

The main issue with this line of reasoning is that both methodologies are inadequate when providing field predictions very close to the interface. This is because the ray tracing formulation completely neglects the existence of interface waves, while the RSI predicts a singularity at the interface, due to the shape of the Green's function used for the sources. To address these issues, we opted to use finite element analysis to obtain the correct wave field solution in these extreme conditions, where other numerical models fail.

\begin{figure}[!b]
    \centering
    \includegraphics[width=1\linewidth]{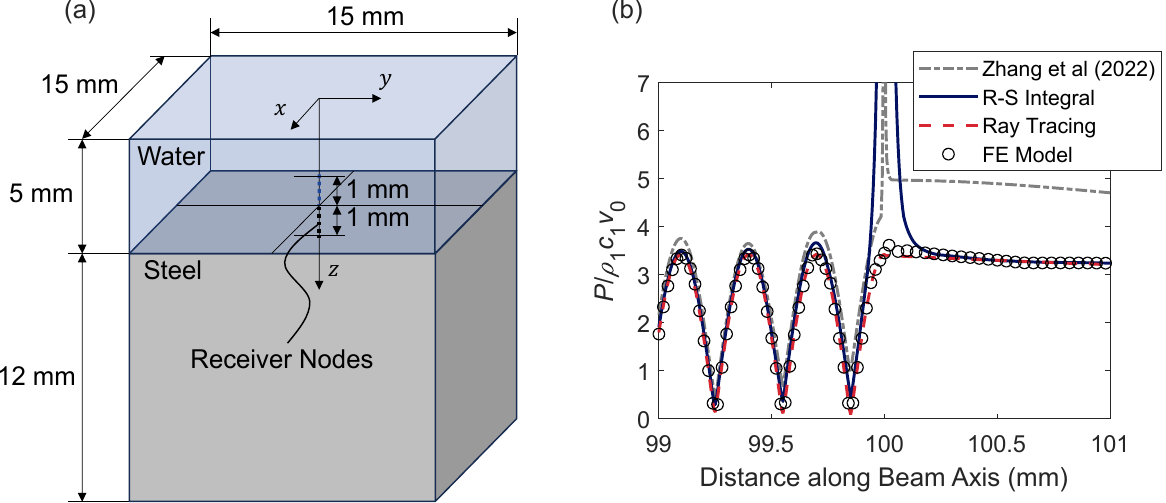}
    \caption{Detailed representation of the geometry used in the FEM (a), and plot of the wave field generated by a transducer at oblique incidence and transmitted through one interface (b). The on-axis amplitude of a dimensionless pressure field is modelled using the RSI and the ray tracing formulation to within $1$ mm on each side of the interface. Results obtained with the model proposed by Zhang et al \cite{Zhang et al_2022} and using FEM are also shown. For the boundary conditions to be satisfied, continuity of pressure should be observed at the interface. The intensity of the field is represented using a normalised pressure amplitude, $P/\rho_1c_1v_o$.}
    \label{fig,boundary_conditions}
\end{figure}

Since we are only interested in the behaviour of the field very close to the boundary, the volume considered can be very small, creating the necessary conditions where FEM can be applied in practice. This constitutes the main reason why the validation using finite elements was not performed for all the other study cases. The model would be too large, either taking several days to reach a solution, or, arguably worse, the size of the displacement vector not allowing its plotting, rendering the results useless.

In fact, as the size of the model increases, so will the number of elements needed to run a FEM of the system, which can be problematic for distances several orders of magnitude larger than the wavelength in that particular medium. The fact that the ray tracing or RSI equations do not change with the size of the system constitutes a major advantage over other element-based numerical techniques, especially for three-dimensional problems involving materials with smaller wavelengths, like fluids.

For the finite element model, we consider a three-dimensional problem with dimensions of 15x15x17 mm. The model is generated using the high-speed GPU-based software Pogo \cite{Huthwaite_2014}, which utilises a central finite differences scheme for time stepping, enabling localised explicit time integration. All outer boundaries are stress-free and the interface between the water and steel is located at $z=5$ mm. On one side of this boundary, between $z=0$ and $z=5$ mm, pressure-based inviscid elements were employed to model non-attenuating longitudinal waves in water. The remaining elements, from $z=5$ mm to $z=17$ mm, were defined as displacement-based elastic elements for solids, which may transmit longitudinal and shear waves. To avoid complications arising from mode conversion at the boundary, all the shear elastic constants of steel were set to 0, effectively imposing a constraint on the possibility of shear wave propagation. Nodes on the xOy plane, at $z=0$ mm, serve as sources, with each particular load function defined as a sinusoidal wave along the z-axis for all time increments. A row of receivers is placed within $1$ mm of the boundary for both media, with their x- and y-coordinates set to $0$.

The particular amplitude and phase of the sinusoidal loading function applied to each source is calculated using our previously established numerical formulations. This strategy is employed, because otherwise the FE model would be too large, if it captured the whole system between the transducer and the interface. The technique used here should not have an impact on our validation attempts, since we are mostly interested in using FE to check how the wave field is transmitted through the interface, as both the RSI and the ray tracing model are well-known to represent accurate wave field propagating in the bulk of a material.

Additionally, the mesh comprises eight-noded brick elements, with a specified element size of $20 \mu$m, equivalent to $30$ elements per longitudinal wavelength. This wavelength is based on its larger value in the medium of water, as opposed to steel. This resolution ensures sufficient detail to accurately capture the wave motion in both media at a centre frequency of $2.25$ MHz, resulting in the model having approximately 1.5 billion Degrees Of Freedom (DOFs). Finally, regarding the temporal evolution of the model, the time step was set to $2.74$ ns to satisfy a stable Courant-Friedrichs-Lewy condition and the runtime was chosen as a suitable compromise between the necessary conditions of our problem. We required a sufficiently large runtime to allow for the emergence of a steady-state solution, while maintaining it small enough such that the field was not subjected to interference with unwanted reflections from the boundaries. The total solution time for this model was two hours, achievable by utilising a High-Performance Computing (HPC) resource equipped with eight Quadro RTX8000 cards. However, this solution time was accompanied by an additional hour of transferring the 44-gigabyte input file to the HPC system.

Figure \ref{fig,boundary_conditions} shows a schematic representation of the FEM geometry and the results obtained with all four methods. It is immediately apparent from the plot, that there is very good agreement between both of our methods and the FE model. This serves as yet another argument in favour of the exactness of the methodologies discussed in this paper, especially when compared to previous work in the literature \cite{Zhang et al_2022}. It is also useful to note that while the FEM required two hours of processing to obtain the information shown in the plot, the RSI-based method and the ray tracing methodology only needed $111$ s and $4$ s respectively.

\subsection{Focused transducer}
\label{sec,focused_trasnducer}
At normal incidence, for the case of a water-steel system being inspected by a focused transducer, the simulation results are plotted in Figure \ref{fig,focused_transducer}. Again, both methodologies are employed to predict a slice of the three-dimensional wave field for $x=0$, the on-axis amplitude, and the off-axis distribution at $z=0.014$ m. The focused transducer used shares the same radius as the planar transducer, albeit with a focal length of $f=0.014$ m into the steel, or conversely a radius of curvature of $R=0.033$ m. Just as in the oblique case, very good agreement is obtained between both methods.

\begin{figure}[!t]
    \centering
    \includegraphics[width=1\linewidth]{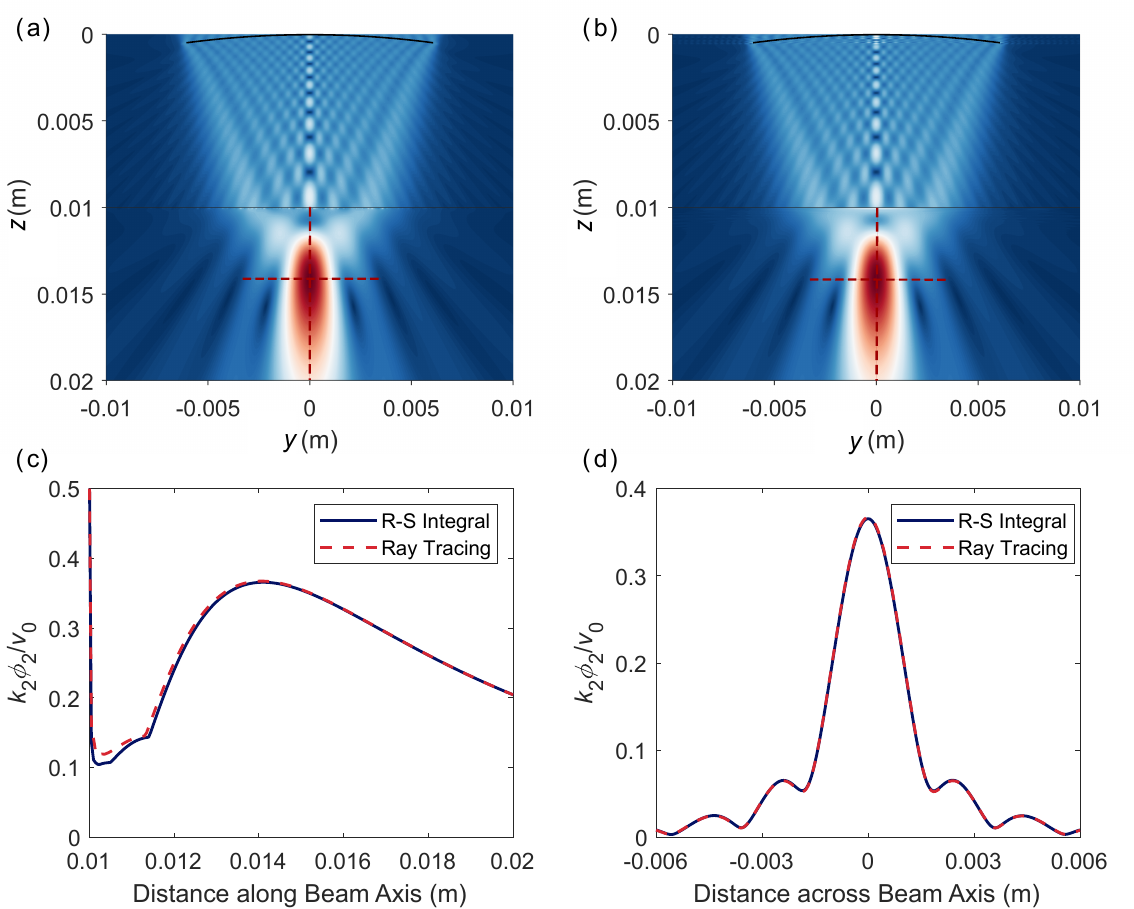}
    \caption{Wave field generated by a focused transducer at normal incidence and transmission through one interface. The full wave field for $x=0$ is modelled using the RSI (a) and the ray tracing formulation (b). Detailed versions of the field are portrayed using the on-axis amplitude of the transmitted field (c) and the off-axis amplitude distribution at $z=0.014$ m (d) for both formulations. The red dashed lines represent the on-axis and off-axis coordinates of the central axis of the beam. The intensity of the field is represented using a normalised velocity amplitude in each medium, $k_i\phi_i/v_o$.}
    \label{fig,focused_transducer}
\end{figure}

\subsection{Multi-layered medium}
\label{sec,layered_media}
Figure \ref{fig,layered_media} shows the results for a 6-layer material immersed in water, with a planar transducer at oblique incidence. The main beam inclination is $10^\circ$ and the layered material is composed of successive layers of steel and aluminium, with $2$ mm of thickness each. Just as in the previous cases, a slice of the three-dimensional wave field for $x=0$, the on-axis amplitude, and the off-axis distribution at $z=0.12$ m are plotted using both formulations.

\begin{figure}[!t]
    \centering
    \includegraphics[width=1\linewidth]{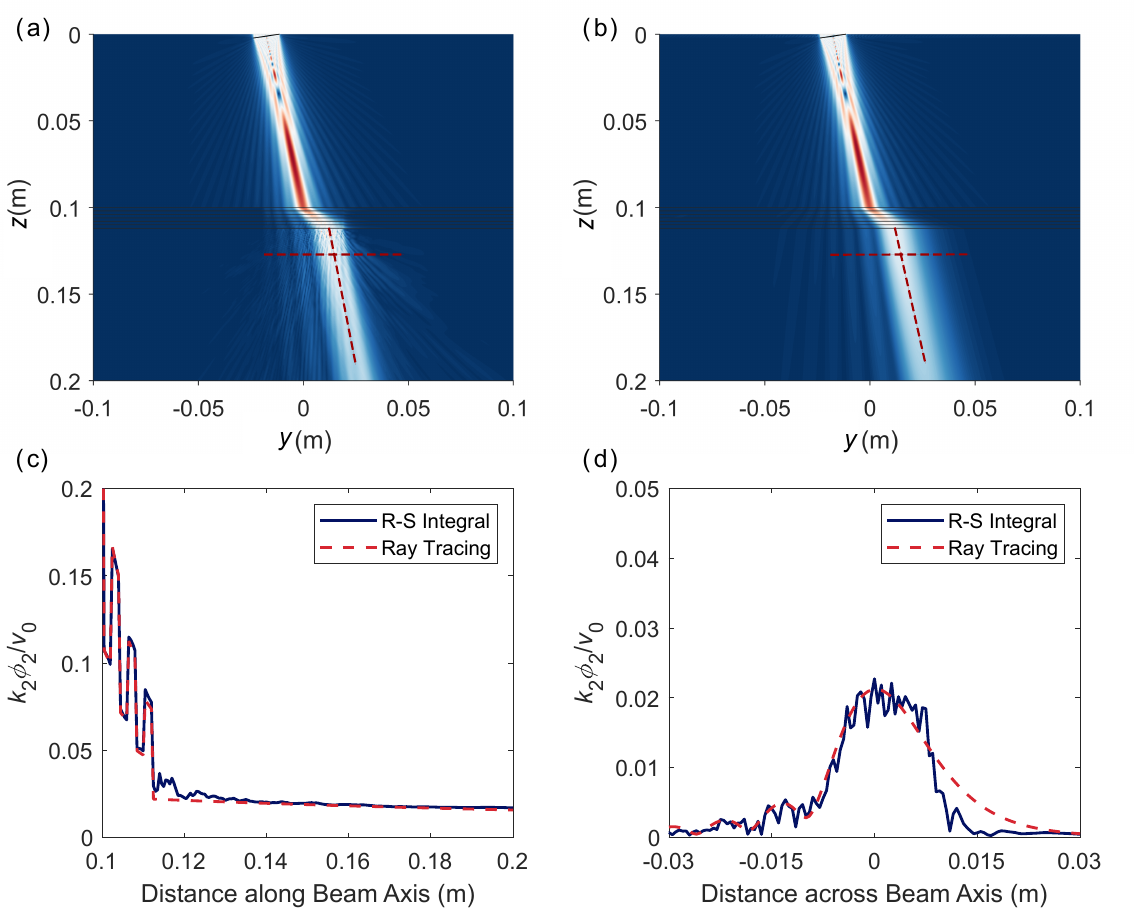}
    \caption{Wave field generated by a transducer at oblique incidence and transmission through eight interfaces. The full wave field for $x=0$ is modelled using the RSI (a) and the ray tracing formulation (b). Detailed versions of the field are portrayed using the on-axis amplitude for $z>0.1$ m (c) and the off-axis amplitude distribution at $z=0.12$ m (d) for both formulations. The red dashed lines represent the on-axis and off-axis coordinates of the central axis of the beam. The intensity of the field is represented using a normalised velocity amplitude in each medium, $k_i\phi_i/v_o$.}
    \label{fig,layered_media}
\end{figure}

Contrary to previous examples, for the layered material the RSI formulation struggled to compute the correct wave field, when compared to the sensible prediction realised by the ray tracing model. The appearance of streaks on the transmitted field, as well as random fluctuations in the on-axis and off-axis profiles, clearly indicate a connection to the quasi-random samples of points used in the QMC integration, which will be explored further in the next section of this paper.

We should note that by increasing the number of sample points in each boundary, the RSI field can be made to approach the real distribution predicted by the ray tracing, at the expense of a substantially larger amount of computational time. However, because we aim to establish a fair comparison between the efficiency of each method, maintaining the same number of sample points was deemed a priority. This also helps highlight one of the main limitations of the RSI model, when it is applied to layered media.

\subsection{Computational efficiency}
\label{sec,computational_efficiency}
One of the key questions we hope to answer with this paper is how to model wave fields generated by transducers effectively and efficiently. It is obvious from our results that a few crucial factors influence the computation time of both methodologies. An increase in the number of field points, of interfaces, or of sample points always lead to longer waiting times, with the relative importance of each of these parameters depending on the formulation being used.

Calculating the field at a point on the other side of an interface is, in general, more complex when using the ray tracing model. This is a direct consequence of the need for root searching, when compared to the purely algebraic computations involved in the RSI formulation. Even if we attempt to use Ferrari's solution for the simple case of a single interface separating two isotropic materials, the amount of necessary computations is large enough that no noticeable improvement to the speed of our algorithm is observed. Therefore, we expect to discern a decrease in efficiency in the ray tracing method as the number of field points increases, when compared to the RSI methodology.

\begin{figure}[!t]
    \centering
    \includegraphics[width=0.7\linewidth]{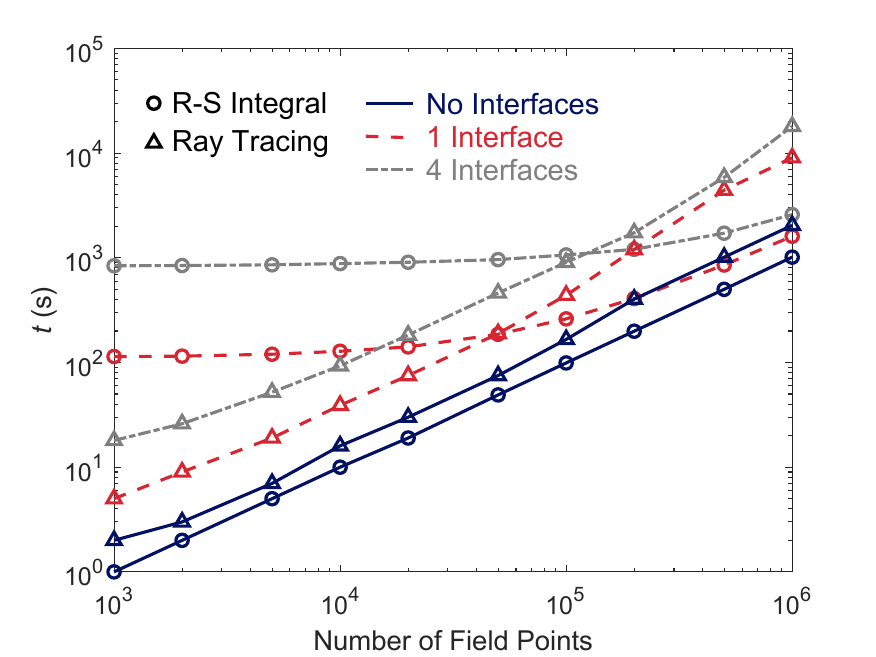}
    \caption{Computation time as a function of the number of field points for several different scenarios using the RSI formulation and the ray tracing model. Three different cases are considered: no interfaces (water), one interface (water-steel), and four interfaces (water-steel-aluminium-steel-water). In all cases, the field points are distributed equally among all the layers.}
    \label{fig,computation_time}
\end{figure}

On the other end of the spectrum, for a small number of field points or a large number of interfaces we anticipate the reverse to be true. Unlike in the ray tracing formulation, where the field is calculated as a direct result from all transducer source points, interfaces constitute computational bottlenecks in the RSI algorithm. This is because to be able to compute the field at a point on the other side of a boundary, the whole field at the interface must be known. This effect is verified in our numerical results, with the ray tracing formulation becoming more efficient for the N-layered example, contrary to what is observed for the single interface examples.

All of these observations are summarised in the plot of Figure \ref{fig,computation_time}. Simulations for three different study cases were carried out using both formulations: a single medium (water) with no interfaces, two media (water-steel) with a boundary in between, and a field with four interfaces (water-steel-aluminium-steel-water). The number of field points was varied with each simulation and the corresponding computation time measured. The number of sample points for the QMC method was kept constant so as to not influence the results obtained from changing the other two parameters, and the field points were distributed equally among all layers. All remaining variables were the same as the ones used in the numerical examples of the previous section.

For a single medium with no interfaces present, we can see that the RSI-based model takes less time to reach the solution, independently of the chosen number of field points. This reiterates our assertion that the ray tracing formulation requires more computations per field point, even without considering the added processing time due to the root searching algorithm. For all other scenarios that include boundaries, the case is not as simple to analyse.

It is possible to see that as the number of field points decreases the computation time for the ray tracing formulation approaches zero unconstrained by the number of interfaces considered. This, of course, is not true for the RSI-based model, as all the curves approach asymptotically a finite constant value, which increases with the number of interfaces. Our statement regarding the interfaces being bottlenecks for this algorithm is thus confirmed, because even if a single point is to be considered, the field needs to be computed at all interfaces.

If the number of field points is increased, it will reach a point where it might be comparable to the number of sample points, summed over all interfaces. From that point onwards, the roles are reversed and it becomes beneficial to use the RSI formulation for evaluating the wave field. The reason for this, as mentioned before, is the increased computational time that a single point calculated using the ray tracing methodology has when compared to an equivalent point using the RSI model. Therefore, if the number of field points becomes very large, the extra time needed to calculate the interface field becomes negligible compared to the time gain for not relying on root searching.

\section{Discussion}
\label{sec,discussion}
From the results shown in the previous section we can conclude that both methodologies provide agreeable predictions, but that their efficient use is highly dependent on the parameters of the system being analysed. We hope to provide a thorough discussion about the scope of each method, their strengths, and limitations.

The RSI model seems a better choice for modelling the field, if a low number of layers is present. This is especially true if the full wave field is required, as the number of points needed is generally large. Besides this, the RSI methodology is simpler to implement, since it does not require binary search or other root finding algorithms as additional complexities, and provides a conceptually cleaner image of the wave field, akin to a more evolved version of the Huygens' construct \cite{Miller_1991}.

As for the ray tracing formulation, it is best suited for applications that do not need the full beam model, but only part of it, so that the number of field points remains low. An example of such an application is the computation of a diffraction correction for a wave travelling between two transducers. In that case, only the field arriving at the receiver is important, possibly making this method faster and more practical.

Another advantage of this method is its ease of application to layered media, without a substantial loss in performance, when compared to simpler systems. This is particularly useful if the objective is to propagate the field through the entire layered material, for example in a through-transmission immersion setup or when inspecting composite materials, provided that an extension to anisotropic media is formulated. In that case, computing the field for a small number of points on a specific layer can be easily achieved without the need for transmitting the field from one interface to the next one, and so on.

On the other hand, the RSI formulation fails to accurately compute the wave field for a layer size of the same order of magnitude as the wavelength. This behaviour arises as a result of the random fluctuations present in the near-field induced by a lack of a sufficiently large sample of interface points. Additionally, because of the steeper asymptotic nature of the dipole, it will be more affected by the near-field effect mentioned when compared to the monopole term. This means that for a layered material characterised by very thin layers, the dipole effect will retain the dominant contribution throughout the whole thickness of each sheet, thereby propagating the random fluctuations, without an observable reduction in amplitude, to every subsequent interface. The only way to fix this issue is by increasing the number of sample points on each interface, which will decrease the algorithm speed further, making the ray tracing formulation much more attractive in comparison.

One other aspect that was not mentioned yet, is the applicability of our methodologies to the propagation of shear waves. This second type of elastic waves is particularly useful after the first critical angle, since for such large angles of incidence, transmitted longitudinal waves vanish, creating a simpler signal to analyse. With the use of a complex transmission coefficient at the boundary, both the correct energy partitioning and the phase shift induced by the evanescent wave on the shear wave can be accurately achieved. Because of this, considering the propagation of shear waves is trivial for the ray tracing case, as opposed to the RSI model, where a different Green's function is needed to model both types of waves present in the solid. This constitutes yet another argument in favour of using ray tracing for anything other than acoustic waves.

Along a similar line, it is relevant to mention the plausible performance of each method when modelling transducer beams propagating through more complex materials. One useful example that comes to mind is the application of our methodologies to the study of anisotropic materials. Without delving into much detail, it seems that the ray tracing formulation might also be better suited for this application, since the only change to the formulation is the need to modulate the slowness according to the direction any specific ray is travelling inside the material \cite{O'Neill and Maev_1998,Every and Amulele_2002}. This is in opposition to the RSI-based model, where, akin to the shear wave case, the Green's functions need to be altered to accommodate the change in wave front shape, relying on the introduction of more nested integrals \cite{Wang and Achenbach_1995}, or on the weak anisotropy approximation \cite{Tverdokhlebov and Rose_1988,Rose et al_1989}.

Other types of applications could also benefit from the use of these numerical methods, such as measuring wave attenuation on thin solid plates or oil film thickness in tribology problems. The latter is known for being quite a complex practical problem to solve, since very small changes in received signals can result in large errors on the predicted thicknesses, which makes it a necessity to rely on finite elements or empirical corrections \cite{Jia et al_2022}. Using our QMC ray tracing methodology could enhance the speed of these calculations, while maintaining the required precision provided by the FEM. For problems with complex geometries, where multiple scatterings might still have a non negligible effect on the final wave field solution, one could instead use a hybrid method that mixes the advantages of both our formulation and of using finite elements. As others have pointed out in the past, one could propagate the field to the scatterer using the QMC ray tracing methodology and subsequently predict the scattered field using a smaller local finite elements simulation \cite{Wilcox and Velichko_2010,Choi et al_2016}.

As for closing remarks, a small discussion regarding the assumptions used for the system excitation is warranted. Firstly, both formulations were derived for a single frequency source, which is of course not representative of actual transducer behaviour. In spite of this, if the complete multi-frequency field is required, both methods can be used to calculate the field for each successive individual frequency, and the compound wave field evaluated using a superposition of all these results. Secondly, the constant intensity profile over the face of the transducer might not be the best approximation for the real distribution, which might more closely resemble a Gaussian. Regardless, it should be straightforward to implement this change using the equations described in this work, by modulating the amplitude of the Green's function according to the position of each specific source in the transducer.

\section{Conclusion}
\label{sec,conclusion}
In this work we implemented two different transducer beam models using a quasi-Monte Carlo integration scheme with the objective of understanding their possible scope and limitations, when modelling wave fields transmitting through interfaces. We also hoped to enhance the RSI-based implementation from a previous paper \cite{Zhang et al_2022}, to offer a more coherent theory for wave field propagation and validate our own ray tracing methodology. Our findings help to contextualise the efficiency, in terms of computation time, of using the pseudo-random generation of sample points to accurately model the field generated by a finite-sized source, as a discretisation technique not attached to any one particular numerical method.

The RSI formulation has the advantage of computing individual field points faster than the ray tracing methodology, since it does not involve any root searching techniques in its algorithm. This makes it an ideal candidate for modelling full field images that require a large number of field points, even when a few interfaces are present.

However, if the goal is to model layered materials with several boundaries, the RSI-based technique struggles to compute an accurate representation of the field, as it needs to propagate the field from each interface to the next one. This makes the method liable to the propagation of random fluctuations, caused by the consideration of quasi-random samples at every interface, which deteriorates the field more and more as we move deeper into the thin-layered material.

Therefore, the ray tracing formulation becomes highly useful to model these types of applications requiring the computation of the wave field inside specific layers of a N-layered material or the transmitted field on the other side of it. This is especially true if a limited number of field points is sufficient for the purpose of a particular application. As a concluding observation, we also believe that for more complex media, such as anisotropic or inhomogeneous materials, the ray tracing formulation will perform better, as its theoretical framework seems more easily adaptable to those cases, when compared to the need for a change in Green's functions for the RSI-based model.

\section{Acknowledgements}
\label{sec,acknowledgements}
The authors would like to gratefully acknowledge funding support from the following sources: the EPSRC Centre for Doctoral Training in Future Innovation in Non-Destructive Evaluation (EP/S023275/1) for André Lello de Almeida; A*STAR Singapore under the National Science Scholarship for Melody Png; the Imperial College Research Fellowship, the NDE group at Imperial College London and the EPSRC grant EP/W014769/1 for Bo Lan. Special thanks are dedicated to Professor Michael J. S. Lowe for revising this paper and providing many helpful insights.


\begin{thebibliography}{}
    \bibitem{Schmerr_2016} Schmerr, L. W. (2016) \textit{Fundamentals of Ultrasonic Nondestructive Evaluation: A Modeling Approach}. 2nd ed. New York, Springer.
    \bibitem{Banerjee et al_2007} Banerjee, S., Kundu, T. \& Alnuaimi, N. A. (2007) DPSM technique for ultrasonic field modelling near fluid-solid interface. \textit{Ultrasonics}. 46, 235-250. doi:10.1016/j.ultras.2007.02.003.
    \bibitem{Zhang et al_2021} Zhang, S., Huang, Y., Li, X. \& Jeong, H. (2021) Modeling of wave fields generated by ultrasonic transducers using a quasi-Monte Carlo method. \textit{J. Acoust. Soc. Am}. 149 (1), 7-15. doi:10.1121/10.0002972.
    \bibitem{Williams_1951} Williams Jr., A. O. (1951) The piston source at high frequencies. \textit{J. Acoust. Soc. Am}. 23, 1-6. doi:10.1121/1.1906722.
    \bibitem{Rogers and Van Buren_1974} Rogers, P. H. \& Van Buren, A. L. (1974) An exact expression for the Lommel diffraction correction integral. \textit{J. Acoust. Soc. Am}. 55 (4), 724-728. doi:10.1121/1.1914589.
    \textit{Review of Progress in Quantitative Nondestructive Evaluation}. Boston, Springer.
    \bibitem{Schmerr et al_1995} Schmerr, L. W., Lerch, T. P. \& Sedov, A. (1995) Modeling the Radiation of Focused and Unfocused Ultrasonic Transducers Through Planar Interfaces. In: Thompson, D. O. \& Chimenti, D. E. (eds.). New York, Plenum Press.
    \bibitem{Lerch et al_1995} Lerch, T. P., Schmerr, L. W. \& Sedov, A. (1995) The Paraxial Approximation for Radiation of a Planar Ultrasonic Transducer at Oblique Incidence Through an Interface. In: Thompson, D. O. \& Chimenti, D. E. (eds.) \textit{Review of Progress in Quantitative Nondestructive Evaluation}. Boston, Springer.
    \bibitem{Wen and Breazeale_1988} Wen, J. J. \& Breazeale, M. A. (1988) A diffraction beam field expressed as the superposition of Gaussian beams. \textit{J. Acoust. Soc. Am}. 83 (5), 1752-1756. doi:10.1121/1.396508.
    \bibitem{Kim et al_2006} Kim, H.-J., Schmerr, L. W. \& Sedov, A. (2006) Generation of the basis sets for multi-Gaussian ultrasonic beam models - An overview. \textit{J. Acoust. Soc. Am}. 119 (4), 1971-1978. doi:10.1121/1.2169921.
    \bibitem{Born and Wolf_1980} Born, M. \& Wolf, E. (1980) \textit{Principles of Optics}. 6th ed. Oxford, Pergamon Press Ltd.
    \bibitem{Goodman_2017} Goodman, J. W. (2017) \textit{Introduction to Fourier Optics}. 4th ed. New York, W. H. Freeman.
    \bibitem{Lord et al_1988} Lord, W., You, Z., Lusk, M. \& Ludwig, R. (1988) Numerical predictions of surface wave phenomena for ultrasonic NDE. In \textit{IEEE 1988 Ultrasonics Symposium Proceedings}. 1065-1068. doi:10.1109/ULTSYM.1988.49541.
    \bibitem{Wojcik et al_1993} Wojcik, G. L., Vaughan, D. K., Abboud, N. \& Mould, J. (1993) Electromechanical modeling using explicit time-domain finite elements. In \textit{IEEE 1993 Ultrasonics Symposium Proceedings}. 1107-1112. doi:10.1109/ULTSYM.1993.339594.
    \bibitem{Bathe_2014} Bathe, K.-J. (2014) \textit{Finite Element Procedures}. 2nd ed. Watertown, MA, Klaus-J\"{u}rgen Bathe.
    \bibitem{Alterman and Karal_1968} Alterman, Z. \& Karal, Jr., F. C. (1968) Propagation of elastic waves in layered media by finite difference methods. \textit{Bull. Seism. Soc. Am}. 58, 367-398. doi:10.1785/BSSA0580010367.
    \bibitem{Scandrett and Achenbach_1987} Scandrett, C. L. \& Achenbach, J. D. (1987) Time-domain finite difference calculations for interaction of an ultrasonic wave with a surface-breaking crack. \textit{Wave Motion}. 9 (2), 171-190. doi:10.1016/0165-2125(87)90051-5.
    \bibitem{Alia et al_2004} Alia, A., Djelouah, H. \& Bouaoua, N. (2004) Finite difference modeling of the ultrasonic field radiated by circular transducers. \textit{J. Comput. Acoust}. 12 (4), 475-499. doi:10.1142/S0218396X04002365.
    \bibitem{Niwa et al_1986} Niwa, Y., Hirose, S. \& Kitahara, M. (1986) Application of the boundary integral equation (BIE) method to transient response analysis of inclusions in a half space. \textit{Wave Motion}. 8 (1), 77-91. doi:10.1016/0165-2125(86)90007-7.
    \bibitem{Aldrin et al_2001} Aldrin, J., Achenbach, J. D., Andrew, G., P'an., C., Grills, B., Mullis, R. T., Spencer, F. W. \& Golis, M. (2001) Case study for the implementation of an automated ultrasonic technique to detect fatigue cracks in aircraft weep holes. \textit{Materials evaluation}. 59 (11), 1313-1319.
    \bibitem{Placko and Kundu_2001} Placko, D. \& Kundu, T. (2001) A theoretical study of magnetic and ultrasonic sensors: dependence of magnetic potential and acoustic pressure on the sensor geometry. Advanced NDE for structural and biological health monitoring. In \textit{Proceedings of SPIE, SPIE's 6th Annual International Symposium on NDE for Health Monitoring and Diagnostics}. 52-62. doi:10.1117/12.434201.
    \bibitem{Lhemery et al_1999} Lh\'{e}mery, A., Paradis, L., Rizzo, L. \& Talvard, M. (1999) Multiple-technique NDT simulations of realistic configurations at the French Atomic Energy Comission (CEA). \textit{Rev. Prog. Quant. Nondestr. Eval}. Vol. 18, 671-678. doi:10.1049/ic:19990107.
    \bibitem{Liu et al_2011} Liu, Y. J., Mukherjee, S., Nishimura, N., Schanz, M., Ye, W., Sutradhar, A., Pan, E., Dumont, N. A., Frangi, A. \& Saez, A. (2011) Recent advances and emerging applications of the boundary element method. \textit{Appl. Mech. Rev}. 64 (3), 030802. doi:10.1115/1.4005491.
    \bibitem{Huthwaite_2014} Huthwaite, P. (2014) Accelerated finite element elastodynamic simulations using the GPU. \textit{J. Comput. Phys. A}. 257, 687-707. doi:10.1016/j.jcp.2013.10.017.
    \bibitem{Zhang et al_2022} Zhang, S., Cheng, C., Li, X., Huang, Y., \& Jeong, H. (2022) Modeling ultrasonic wave fields using a Quasi-Monte Carlo method: Wave transmission through complicated interfaces. \textit{J. Acoust. Soc. Am}. 152 (2), 994-1002. doi:10.1121/10.0013411.
    \bibitem{Rayleigh_1945} Rayleigh, J. W. S. (1945) \textit{The Theory of Sound}, Vol. II. New York, Dover.
    \bibitem{King_1934} King, L. V. (1934) \textit{Can. J. Res}. 11, 135-146.
    \bibitem{Schoch_1941} Schoch, A. (1941) \textit{Akust. Z}. 6, 318-326.
    \bibitem{Harris_1981} Harris, G. R. (1981) Review of transient field theory for a baffled planar piston. \textit{J. Acoust. Soc. Am}. 70 (1), 10-20. doi:10.1121/1.386687.
    \bibitem{Nayfeh and Chimenti_1984} Nayfeh, A. H. \& Chimenti, D. E. (1984) Reflection of finite acoustic beams from loaded and stiffened half-spaces. \textit{J. Acoust. Soc. Am}. 75 (5), 1360-1368. doi:10.1121/1.390857.
    \bibitem{Pialucha_1992} Pialucha, T. P. (1992) \textit{The reflection coefficient from interface layers in NDT of adhesive joints}. PhD thesis. Imperial College London.
    \bibitem{Rubinowicz_1917} Rubinowicz, A. (1917) \textit{Ann. Physik}. 53, 257.
    \bibitem{Marchand and Wolf_1962} Marchand, E. W. \& Wolf, E. (1962) Boundary diffraction wave in the domain of the Rayleigh-Kirchhoff diffraction theory. \textit{J. Opt. Soc. Am}. 52 (7), 761-767. doi:10.1364/JOSA.52.000761.
    \bibitem{Miller_1991} Miller, D. A. B. (1991) Huygens's wave propagation principle corrected. \textit{Opt. Lett}. 16 (18), 1370-1372. doi:10.1364/OL.16.001370.
    \bibitem{Seki et al_1956} Seki, H., Granato, A. \& Truell, R. (1956) Diffraction Effects in the Ultrasonic Field of a Piston Source and Their Importance in the Accurate Measurement of Attenuation. \textit{J. Acoust. Soc. Am}. 28 (2), 230-238. doi:10.1121/1.1908249.
    \bibitem{Weston et al_2012} Weston, M., Mudge, P., Davis, C. \& Peyton, A. (2012) Time efficient auto-focussing algorithms for ultrasonic inspection of dual-layered media using Full Matrix Capture. \textit{NDT\&E International}. 47, 43-50. doi:10.1016/j.ndteint.2011.10.006.
    \bibitem{Morokoff and Caflisch_1995} Morokoff, W. J. \& Caflisch, R. E. (1995) Quasi-Monte Carlo Integration. \textit{J. Comput. Phys}. 122, 218-230. doi:10.1006/jcph.1995.1209.
    \bibitem{O'Neill and Maev_1998} O'Neill, B. E. \& Maev, R. G. (1998) Integral approximation method for calculating ultrasonic beam propagation in anisotropic materials. \textit{Phys. Rev. B}. 58, 5479-5485. doi:10.1103/PhysRevB.58.5479.
    \bibitem{Every and Amulele_2002} Every, A. G. \& Amulele, G. M. (2002) Angular spectrum method and ray tracing algorithm for the acoustic field of a focusing transducer in an anisotropic solid. \textit{IEEE Transactions on Ultrasonics, Ferroelectrics, and Frequency Control}. 49 (3), 307-318. doi:10.1109/58.990942.
    \bibitem{Wang and Achenbach_1995} Wang, C.-Y. \& Achenbach, J. D. (1995) Three-dimensional time-harmonic elastodynamic Green's functions for anisotropic solids. \textit{Proc. R. Soc. Lond. A}. 449, 441-458. doi:10.1098/rspa.1995.0052.
    \bibitem{Tverdokhlebov and Rose_1988} Tverdokhlebov, A. \& Rose, J. (1988) On Green's functions for elastic waves in anisotropic media. \textit{J. Acoust. Soc. Am}. 83, 118-121. doi:10.1121/1.396437.
    \bibitem{Rose et al_1989} Rose, J., Balasubramaniam, K. \& Tverdokhlebov, A. (1989) A numerical integration Green's function model for ultrasonic field profiles in mildly anisotropic media. \textit{J. Nondestruct. Eval}. 8, 165-179. doi:10.1007/BF00570885.
    \bibitem{Jia et al_2022} Jia, Y., Dou, P., Zheng, P., Wu, T., Yang, P., Yu, M. \& Reddyhoff, T. (2022) High-accuracy ultrasonic method for in-situ monitoring of oil film thickness in a thrust bearing. \textit{Mechanical Systems and Signal Processing}. 180. doi.org/10.1016/j.ymssp.2022.109453.
    \bibitem{Wilcox and Velichko_2010} Wilcox, P. D. \& Velichko, A. (2010) Efficient finite element modeling of scattering for 2D and 3D problems. \textit{Proc. SPIE 7650, Health Monitoring of Structural and Biological Systems 2010}. 76501E. doi.org/10.1117/12.847569.
    \bibitem{Choi et al_2016} Choi, W., Skelton, E. A., Pettit, J., Lowe, M. J. S. \& Craster, R. V. (2016) A generic hybrid model for the simulation of three-dimensional bulk elastodynamics for use in nondestructive evaluation. \textit{IEEE Transactions on Ultrasonics, Ferroelectrics, and Frequency Control}. 63 (5), 726-736. doi.org/10.1109/TUFFC.2016.2535369.
\end{thebibliography}
\end{document}